\documentclass[conference, a4, 10pts]{IEEEtran}
\usepackage{cite}
\usepackage[pdftex]{graphicx}
\DeclareGraphicsExtensions{.pdf,.jpeg,.png}
\usepackage{amsmath}
\usepackage{algorithmic}
\usepackage{array}
\usepackage{fixltx2e}
\usepackage{url}
\usepackage{color,soul}
\usepackage{caption}
\usepackage{subcaption}
\usepackage{multirow}
\usepackage{authblk}
\IEEEoverridecommandlockouts
\begin{document}
\title{Characterizing the Communication Requirements of GNN Accelerators: A Model-Based Approach}

\author{Robert Guirado}
\author{Akshay Jain}
\author{Sergi Abadal}
\author{Eduard Alarc\'on\\[-4.0ex]}
\affil{Universitat Polit\`ecnica de Catalunya}


\maketitle

\begin{abstract}
Relational data present in real world graph representations demands for tools capable to study it accurately. In this regard Graph Neural Network (GNN) is a powerful tool, wherein various models for it have also been developed over the past decade. Recently, there has been a significant push towards creating accelerators that speed up the inference and training process of GNNs. These accelerators, however, do not delve into the impact of their dataflows on the overall data movement and, hence, on the communication requirements. In this paper, we formulate analytical models that capture the amount of data movement in the most recent GNN accelerator frameworks. Specifically, the proposed models capture the dataflows and hardware setup of these accelerator designs and expose their scalability characteristics for a set of hardware, GNN model and input graph parameters. Additionally, the proposed approach provides means for the comparative analysis of the vastly different GNN accelerators.
\end{abstract}

\begin{IEEEkeywords}
GNN Accelerators, Parametric models
\vspace{-0.3cm}
\end{IEEEkeywords}

\section{Introduction} \label{introd}
\vspace{-0.1cm}
Representation of data-sets as graph-based data structures, given the opportunities they provision to understand complex relationships embedded in them, has become increasingly popular \cite{Bronstein2017, Battaglia2018, Zhang2020a,eisen2020optimal, abadal2020computing}. These graph representations can range from very small (chemistry) to extremely huge (recommendation systems) graphs \cite{Fan2019,Fout2017}. Subsequently, to learn and infer from these graph representations, the seminal work by Scarselli \emph{et al.} \cite{Scarselli2009} in 2009 introduced the notion of Graph Neural Networks (GNNs). 
While multiple GNN models have been developed over the past decade \cite{Kipf2016, Kipf2019, Bresson2018, ying2018hierarchical,Dwivedi2017,Hamilton2017a,Schlichtkrull2018}, it is only recently that accelerators for speeding up the training and inference of GNNs have gained attention \cite{liang2020engn,Yan2020,AWB,Auten2020,Kiningham2020,Zhang2020acc,Zeng2020}. 


The area of GNN acceleration, however, is still in its infancy and the handful of accelerators that exist cover a wide portion of the design space, but only with few sparse points %
\cite{abadal2020computing}. Moreover, as shown in Table \ref{tab1}, these accelerators support a broad variety of GNN algorithm variants. While the accelerator designs verify their performance through metrics such as throughput and overall energy consumption, an explicit study on the data movement within the accelerators is absent. Note that, the data movement dictates the requirements cast on the underlying on-chip interconnect, which has a large impact on the latency and energy consumed by the accelerator. 
In fact, an important aspect of current GNN accelerator designs is to minimize data movement in order to be faster, more scalable, and more energy efficient. Hence, there is an interest on studying the effect of the dataflows and hardware design of the accelerators on the scalability trends of their underlying interconnect fabric \cite{understanding}. This will also provision important insights which will help guide future GNN accelerator designs. 


\begin{table}[!htb]    
\centering
    \renewcommand{\arraystretch}{0.9}
    \caption{Selection of GNN accelerators in the literature.}
    \vspace{-0.2cm}
    \begin{tabular}{|m{1.9cm}|m{6.1cm}|} \hline
        \textbf{Accelerators} & \textbf{Algorithms Supported}\\ \hline
        EnGN \cite{liang2020engn} & GCN \cite{Kipf2016}, GraphSage-Max \cite{Hamilton2017a}, GatedGCN \cite{Bresson2018}, GRN \cite{liang2020engn}, R-GCN \cite{Schlichtkrull2018} \\ \hline
        HyGCN \cite{Yan2020} & GCN, GraphSage-Mean \cite{Hamilton2017a}, GIN \cite{Xu2019}, DiffPool \cite{ying2018hierarchical} \\ \hline
        Auten \emph{et al.} \cite{Auten2020} & GCN, GAT \cite{Dwivedi2017}, PGNN \cite{Auten2020} \\ \hline
        AWB-GCN \cite{AWB} & GCN \\ \hline
        GRIP \cite{Kiningham2020} & GCN, GraphSage-Max, GIN, GatedGCN \\ \hline
        
    \end{tabular}
    \label{tab1}
    \vspace{-0.1cm}
\end{table}

\noindent In this context, this paper makes the following contributions:  

\begin{itemize}
    \item It presents analytical models that describe the impact of the dataflows and hardware design of GNN accelerators on the overall data movement. The models are based on the accurate descriptions of the accelerator in question. To the best of our knowledge, this is the first work providing such analytical models, which will help in performing scalability analyses, comparing between different accelerator designs, and making more informed decisions with regards to future GNN accelerators. 
    \item It evaluates two distinct state-of-the-art GNN accelerators, i.e., EnGN \cite{liang2020engn} and HyGCN \cite{Yan2020}, using the proposed models. This opens the door to the development of similar analytical models for other accelerators. 
\end{itemize}

We provide background on GNNs in Sec. \ref{bcg}, describe the models in Sec. \ref{analys}, discuss the scaling results in Sec. \ref{results}, and conclude the paper in Sec. \ref{cncl}.


\section{GNN Accelerators: Background}\label{bcg}
GNN accelerators are application-specific integrated circuits that aim to process one or multiple GNN variants  (see Table \ref{tab1}) in a timely and energy efficient manner. The main challenge in GNN accelerator design is the alternation of phases with either dense or extremely sparse computation. The sparsity is driven by the graph connectivity or the graph adjacency matrix \cite{liang2020engn,Yan2020,AWB}. On the other hand, phases of dense computations are usually due to the dense nature of the operations that are applied to the nodes and edges in parallel \cite{liang2020engn,Yan2020,AWB}. Additionally, GNNs process input graphs that might have billions of nodes and edges, with uneven connectivity. Such characteristics lead to workload profiles that differ from conventional neural networks (NNs), besides being inherently imbalanced \cite{AWB}.

\begin{figure} [!htb]
    \centering
    \vspace{-0.4cm}
    \includegraphics[height=3cm,width=\columnwidth]{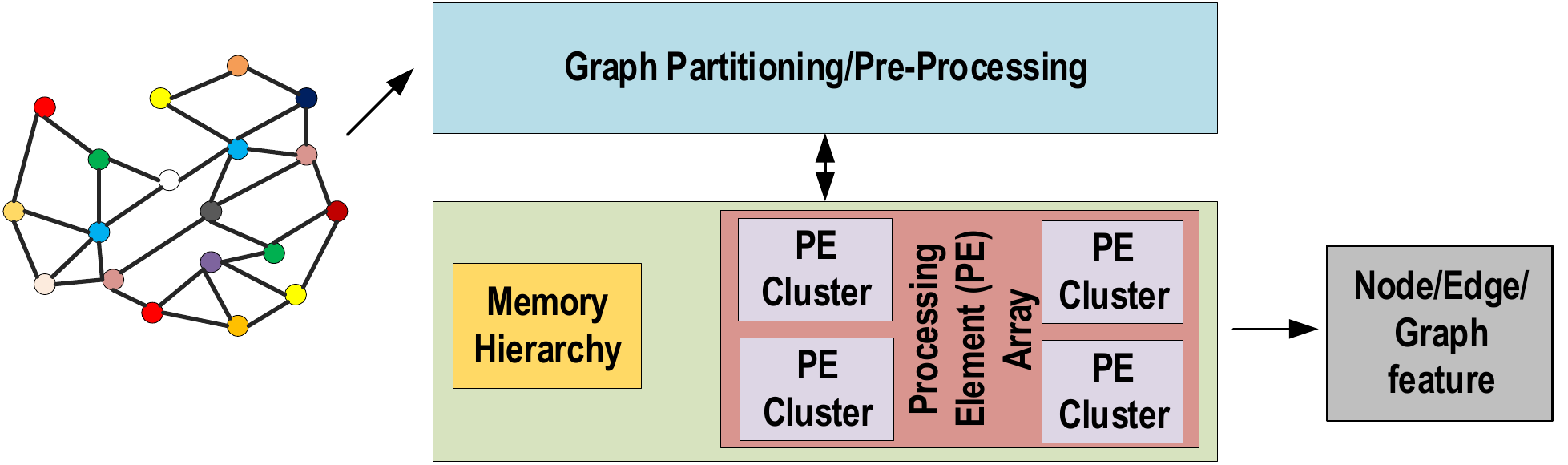}
    \caption{Generic schematic diagram of a GNN accelerator.}
    
    \label{GNNAccel}
    \vspace{-0.3cm}
\end{figure}

Consequently, multiple works \cite{liang2020engn,Yan2020,AWB,Auten2020,Kiningham2020,Zhang2020acc,Zeng2020} propose accelerators that speedup the inference/learning process with various specific architectural techniques. To illustrate the fundamental principles, Fig. \ref{GNNAccel} represents a generic architecture of a GNN accelerator with its corresponding inputs and outputs. Concretely, the acceleration engine takes as input a graph data structure and performs some initial pre-processing, such as tiling/graph partitioning \cite{liang2020engn}. This is then fed to the processing engine, which depending on its architecture and the predefined dataflow, processes the incoming graph partitions to yield an output. This output is generally a vector of features (i.e. predictions) for either a node, an edge, or the entire graph.       

The GNN inference process is generally divided into two computation stages, namely, aggregation and combination \cite{abadal2020computing}. First, the data from nodes or edges is loaded from the memory hierarchy to the Processing Elements (PEs). In the aggregation stage, the node and/or edge features are added to that of their neighbours following the adjacency matrix, resulting into movement of data across PEs. Subsequently, in the combination stage, the aggregated features are transformed through a set of typical NN operations such as matrix-vector multiplications and a non-linear activation function (e.g., ReLU or tanh). Finally, after multiple recursions of the aforesaid stages, a readout of the desired features is performed via another NN or by just a linear/non-linear transformation. Training follows a similar process \cite{abadal2020computing}.

\begin{figure} [!htb]
    \centering
    \begin{subfigure}{0.45\columnwidth}
    \includegraphics[height=3cm, width=\columnwidth]{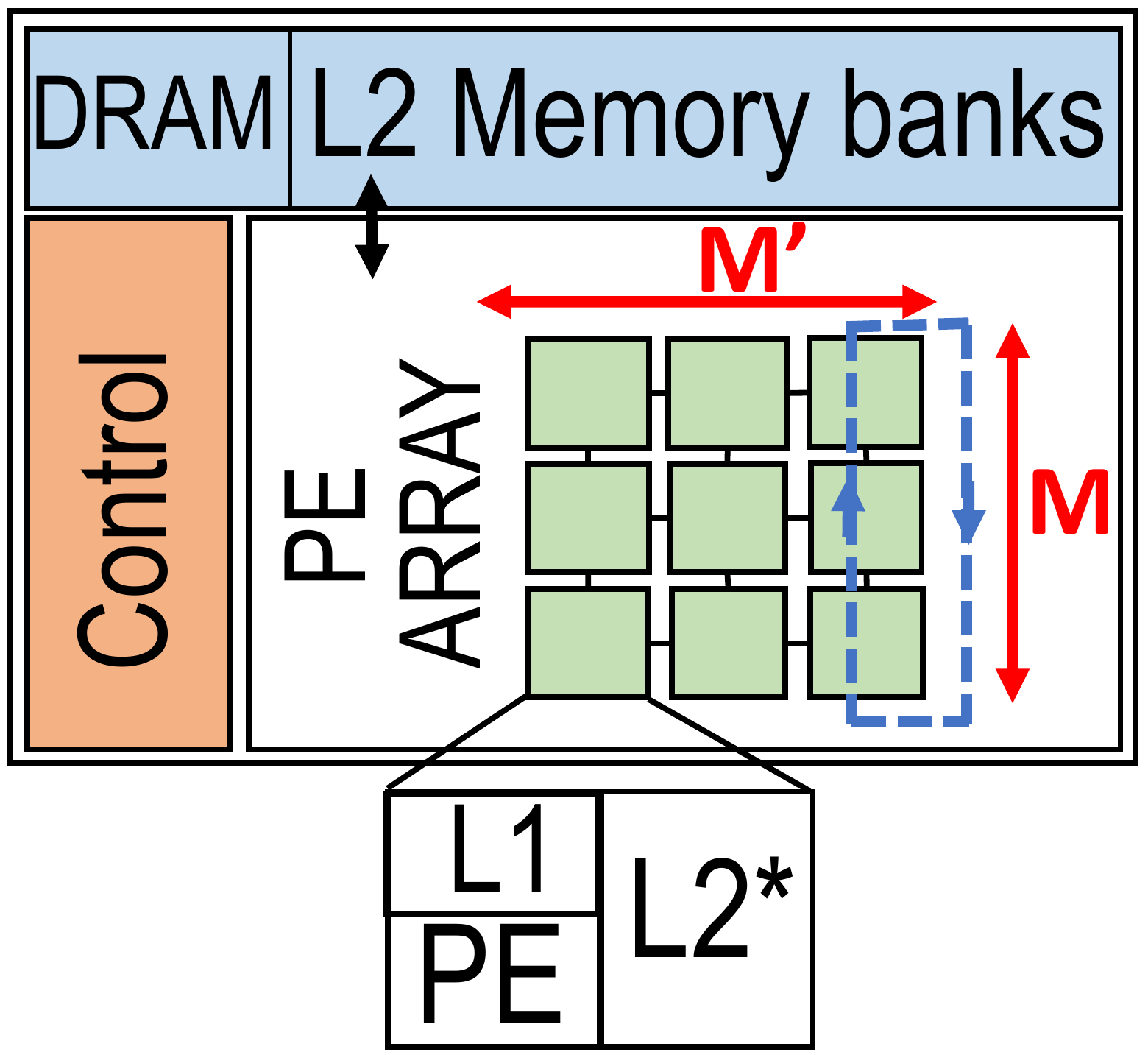}
    \caption{EnGN \cite{liang2020engn}}
    \end{subfigure}
    \begin{subfigure}{0.45\columnwidth}
    \includegraphics[height=3cm, width=4.5cm]{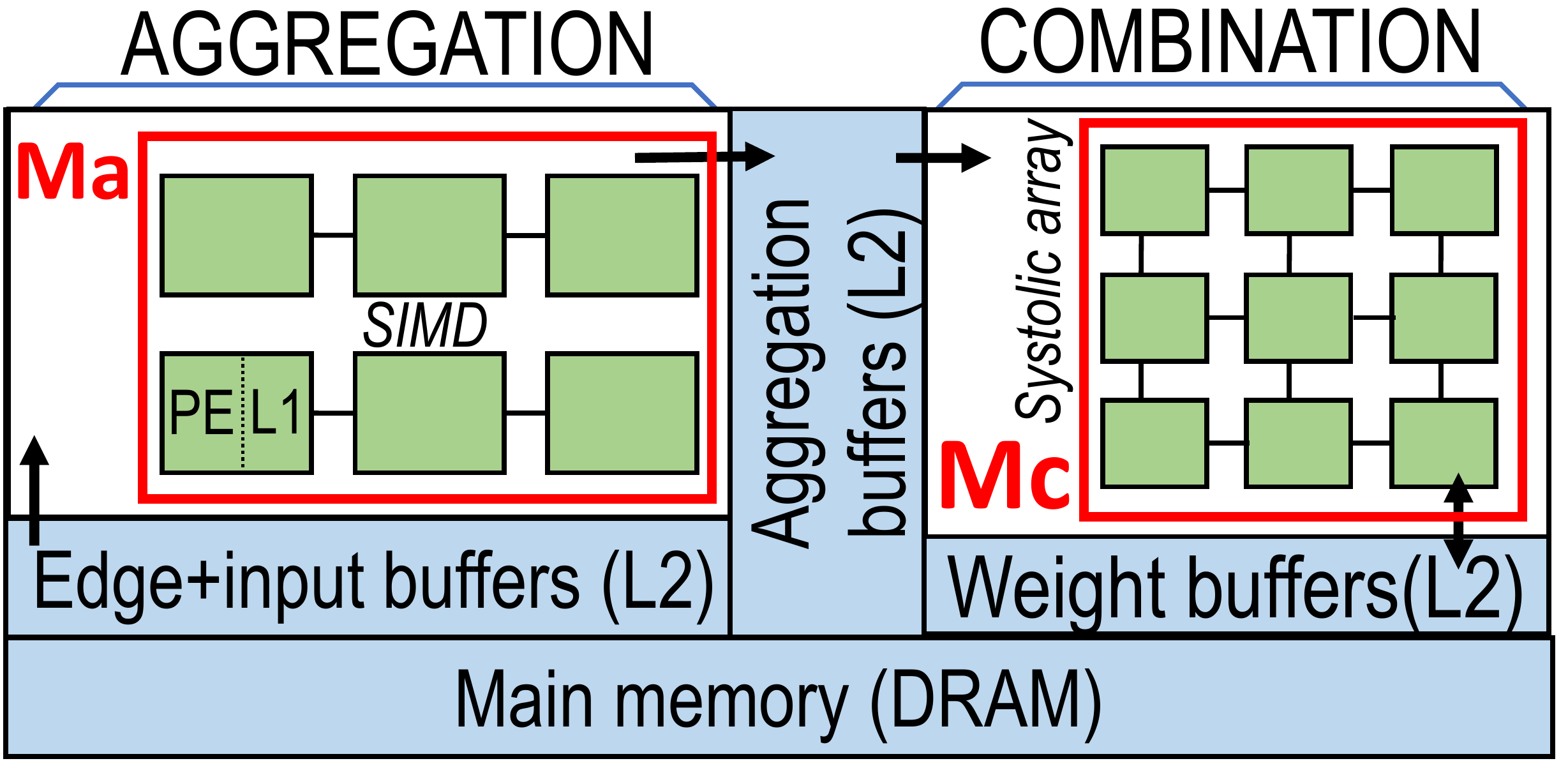}
    \caption{HyGCN \cite{Yan2020}}
    \end{subfigure}
    \vspace{-0.1cm}

    \caption{Schematic diagram of the analyzed GNN accelerators.} 
    \vspace{-0.4cm}
    \label{fig:acc_schemes}
\end{figure}

Given their distinct architectures, in this work we focus on two state-of-the-art GNN acceleration engines, i.e., EnGN \cite{liang2020engn} and HyGCN \cite{Yan2020}. Figs. \ref{fig:acc_schemes}(a) and \ref{fig:acc_schemes}(b) illustrate their hardware architecture, 
which implement vastly differing methods for, among others, the graph partitioning and the reuse of vertex features and intermediate results. These design decisions, along with the specific dataflow they implement, impact the characteristics of data movement. Hence, we next describe analytical models of data movement for processing a single tile, that can help understand these characteristics.

\section{Analytical Models for Data Movement} \label{analys}
In this section, we describe the process towards obtaining the parametric models that characterize data movement in GNN accelerators. Table \ref{tab:notation} lists the employed notation. In essence, we consider different input graph parameters, such as the number of nodes $K$ and edges $P$ in a tile, and different hardware parameters, such as the amount of differently-organized PEs $M, M', M_{a}, M_{c}$ and memory bandwidth $B$ [bits/iteration], to breakdown the necessary data movement in a given accelerator based on its defined dataflow. 

The models determine the \emph{amount of data movement}, this is, the total number of bits that must be moved between different memory hierarchy levels; and the \emph{number of iterations}, this is, the number of steps required to move all that data due to PE or memory bandwidth constraints. High data movement will likely lead to high energy consumption, while many iterations may be an indicator of high latency. The hierarchy levels are also specified as they have an obvious impact on latency and energy. For instance, accessing 
a memory bank (L2) is
$\sim$6$\times$ more expensive than accessing a register file (L1) \cite{eyeriss}.


The proposed models are based on the analysis of the thorough architectural descriptions and walkthrough examples of EnGN \cite{liang2020engn} and HyGCN \cite{Yan2020}. Validation of the data movement models is difficult as the authors of both accelerators provide metrics normalized to the CPU or GPU implementations mostly, and do not explicitly study data movement. Moreover, their simulation tools are in-house and not open source.

\begin{table}[!htb]    
\centering
\setlength\tabcolsep{3pt}
    \renewcommand{\arraystretch}{0.9}
    \vspace{-0.2cm}
    \caption{Summary of notation.}
    \vspace{-0.2cm}
    
    \begin{tabular}{|>{\centering\arraybackslash}m{0.3cm}|>{\centering\arraybackslash}m{3.4cm}||>{\centering\arraybackslash}m{0.8cm}|>{\centering\arraybackslash}m{3.4cm}|}
    \hline
    \multicolumn{2}{|c||}{\textbf{Input graph parameters}} &  \multicolumn{2}{c|}{\textbf{Architecture parameters}}\\
    \hline
    N & Size of input feature vector & $\sigma$ & Bit precision \\
    \hline
    T & Size of output feature vector & B & L2 memory bandwidth \\
    \hline
    K & Number of vertices in a tile &  M$\times$M' & EnGN PE array size \\ \hline
    L & Number of high-degree vertices in a tile & M\textsubscript{a} M\textsubscript{c} & HyGCN aggregation and combination PEs \\ \hline  
    \multirow{2}{*}{P} & \multirow{2}{*}{Number of edges in a tile} & $\Gamma$ & HyGCN systolic array reuse\\     \cline{3-4}
     &  & P\textsubscript{s} & HyGCN edges after sliding \\
    \hline
    
    \end{tabular}
    \label{tab:notation}
\end{table}

\begin{table*}[!htb]    
\centering
    \renewcommand{\arraystretch}{0.9}
    \caption{EnGN analytical model. L2* refers to a dedicated vertices cache.}
    \vspace{-0.2cm}
    \begin{tabular}{|c|c|c|c|}
    \hline
    \textbf{Movement level} & \textbf{Data movement} & \textbf{Number of iterations} & \textbf{Hierarchy} \\
    \hline
    loadvertcache & min(L$\sigma$,M$\sigma$,B*)$\cdot$N$\cdot$ceil(L$\sigma$/min(B*,M$\sigma$)) & ceil(L$\sigma$/min(B*,M$\sigma$)) & L2*--L1 \\
    \hline
    loadvertL2 & min((K–L)$\sigma$,M$\sigma$,B)$\cdot$N$\cdot$ceil((K-L)$\sigma$/min(B,M$\sigma$)) & ceil((K-L)$\sigma$/min(B,M$\sigma$)) & L2--L1 \\
    \hline
    loadedges & min(P$\sigma$,B)$\cdot$ceil(P$\sigma$/B) & ceil(P$\sigma$/B) & L2--L1 \\
    \hline
    loadweights & min(T$\sigma$,M$\sigma$,B)$\cdot$N$\cdot$ceil(T$\sigma$/min(B,M$\sigma$)) & ceil(T$\sigma$/min(B,M$\sigma$)) & L2--L1  \\
    \hline
    aggregate & M(M-1)T$\cdot$(ceil(K/M)+ceil(K(N-M)/M))$\sigma$ & ceil(K/M)+ceil(K(N-M)/M) & L1--L1  \\
    \hline 
    writecache & min(M$\sigma$,L$\sigma$, B*)T$\cdot$ceil(L$\sigma$/min(M$\sigma$,B*)) & ceil(L$\sigma$/min(M$\sigma$,B*)) & L1--L2  \\
 \hline
    writeL2 & min(M$\sigma$,(K-L)$\sigma$, B)T$\cdot$ceil((K-L)$\sigma$/min(M$\sigma$,B)) & ceil((K-L)$\sigma$/min(M$\sigma$,B)) & L1--L2  \\
 \hline
     
    \end{tabular}
    \label{tab:EnGN_model}
    \vspace{-0.2cm}
\end{table*}

\begin{table*}[!htb]    
\centering
    \renewcommand{\arraystretch}{0.9}
    \caption{HyGCN analytical model.}
    \vspace{-0.2cm}
    \begin{tabular}{|c|c|c|c|}
    \hline
    \textbf{Movement level} & \textbf{Data movement} & \textbf{Number of iterations} & \textbf{Hierarchy} \\
    \hline
    loadvertL2 & min(K$\sigma$,M$_a \sigma$,B))$\cdot$N$\cdot$ceil(K$\sigma$/min(B,M$_a \sigma$) & ceil(K$\sigma$/min(B,M$_a\sigma$)) & L2--L1  \\
    \hline
    loadedges & min(P$_s\sigma$,B)ceil(P$_s\sigma$/B) & ceil(P$_s\sigma$/B) & L2--L1 \\
    \hline
    loadweights & min(NT$\sigma(1-\Gamma)$,M$_{c}\sigma$,B)ceil(NT$\sigma(1-\Gamma)$/min(B, M${_c}\sigma$)) & ceil(NT$\sigma(1-\Gamma)$/min(B, M${_c}\sigma$)) & L2--L1 \\
    \hline
    aggregate & min(NP$_s\sigma$,M$_a$8)ceil(NP$_s\sigma$/(M$_a$8))  &ceil(NP$_s\sigma$/(M$_a$8)) & L1--L1 \\
    \hline
    writeinterphase & min(KN$\sigma$, B)ceil(KN$\sigma$/B)
 &ceil(KN$\sigma$/B) & L1--L2 \\
    \hline
    combine & KN$\sigma+$ NT$\sigma$ & 1 & L1--L1 \\
    \hline
    readinterphase & min(P$_s$N$\sigma$,B,M$_c$)ceil(P$_s$N$\sigma$/min(B,M$_c$))
 & ceil(P$_s$N$\sigma$/min(B,M$_c$)) & L2--L1 \\
    \hline
    writeL2 & min(KT$\sigma$, B)ceil(KT$\sigma$/B) & ceil(KT$\sigma$/B) & L1--L2\\
    \hline
    \end{tabular}
    \label{tab:HyGCN_model}
    \vspace{-0.4cm}
\end{table*}

Table \ref{tab:EnGN_model} lists the amount of data movement and number of iterations in the different stages of EnGN. 
By default, this accelerator employs a single $128\times 16$ PE array alongside multiple levels of memory to process both the aggregation and combination stages sequentially. First, a control engine determines how the input graph is streamed onto the processing engine. Vertices, both from L2 and a cache storing highly-connected vertices, start being loaded as stated in \textit{loadvertL2} and \textit{loadvertcache} respectively. We consider limitations in memory bandwidth $B$, and row PE array size $M$. Next, for the aggregation stage, a ring-edge-reduce (RER) whereby PEs send their outputs following a physical ring is proposed \cite{liang2020engn}. RER generates the data movement specified in \textit{aggregate} due to its ring nature. After this, EnGN loads the weights for the combination stage, \textit{loadweights}, we take into account the required amount of weights to process a tile, i.e., $N\times T$. Further, to compute the data that can be read in each iteration, we consider the minimum between $T$, the memory bandwidth $B$ and the compute array size. We chose the minimum of the aforesaid parameters since in the EnGN architecture vertex features are loaded on the PE engine in a streamed fashion. Finally, results are written in the cache and the L2 memory bank, \textit{writecache} and \textit{writeL2}, and the next tile is loaded, in \textit{intertile}. The required iterations, as mentioned above, are then the result of a ceiling function between the total data movement and the data that can be moved at once. 


Table \ref{tab:HyGCN_model}, on the other hand, presents the model for HyGCN. HyGCN consists of two separate engines for the pipelined execution of the aggregation (PE array with 32 Single-Input-Multiple-Data cores) and combination (a systolic array with $8 \times 4 \times 128$ PEs). The memory organization is slightly different than in EnGN, using an extra aggregation buffer to store intermediate results \cite{Yan2020}. 
To account for these particularities, we introduce a $\Gamma$ parameter, which reflects the reusability of data within the combination engine, i.e., the systolic array. To process a tile in HyGCN, graph data is loaded from L2 input and edge buffers, \textit{loadvertL2} and \textit{loadedges}. The aggregation task, \textit{aggregation}, is handled by all vertices at the same time, each of them working on up to eight separate feature components. Then, aggregated features are stored in the inter-phase cache, ready to be fetched by the combination engine, \textit{writeinterphase} and \textit{readinterphase}, and proceed with the matrix-vector multiplications. The only remaining step is to write the results to the output buffer, \textit{writeL2}.
\vspace{-0.2cm}




\section{Results and Discussion} \label{results}


We evaluate the total data movement, in bits, and number of iterations for the processing of a tile, i.e., fixed-size portion of a graph. One could extend the analysis to arbitrary graphs by multiplying by its number of tiles and modeling the reuse across tiles. 
Here, we sweep both input (number of vertices $K$) and architectural parameters (number of PEs and bandwidth) to obtain scalability trends. Other parameters have default values of $N=30$, $T=5$, $B=1000$, $\sigma=4$. The number of edges is $P=10\cdot K$ to model similar connectivity in growing tiles. We set $P_{s} \sim P$ as the performance of HyGCN's sliding window may vary across tiles.


\subsection{EnGN}

\begin{figure*}[!htb]
    \centering
    \includegraphics[width=\textwidth]{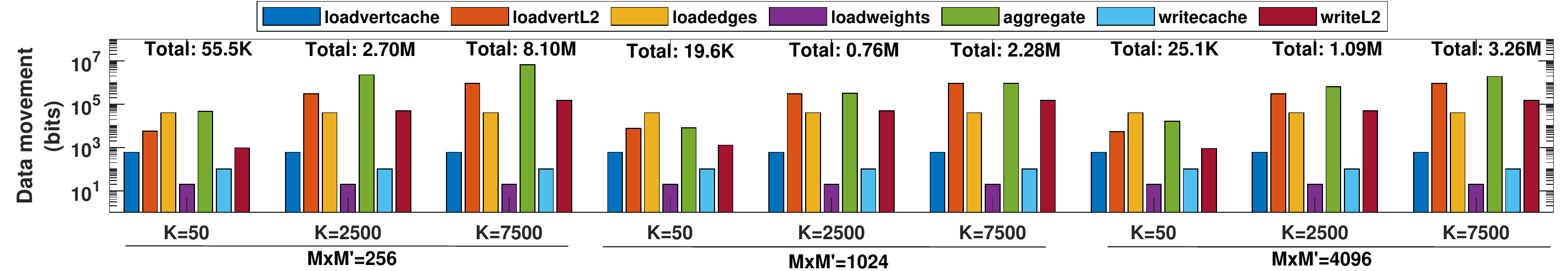}
    \vspace{-0.7cm}
    \caption{EnGN amount of data movement with varying input graph and accelerator configuration.} 
     \vspace{-0.4cm}
    \label{fig:EnGN_bar}
\end{figure*}

Fig. \ref{fig:EnGN_bar} shows the data movement in an EnGN-like accelerator, for different graph sizes and PE array size. We take $M=M'$ for the sake of clarity. It can be observed how aggregation dominates and leads to over 10$\times$ more data movement than \textit{loadvertL2}. This is due to the RER-based aggregation strategy. However, since this movement is happening between L1 memory levels, it is faster and less energy-consuming than other data paths. Moreover, we can observe how data movement increases linearly with $K$, but not with $M\times M'$. In the latter case, small PE arrays struggle to compute the tile efficiently as they need to continuously fetch new values into the RER, which increases data movement. Another interesting result is that \textit{loadvertcache} data, fetched from the dedicated cache for highly-connected vertices, relieves significant load from the vertex memory bank at \textit{loadvertL2}. As compared to HyGCN, EnGN shows a smaller \textit{loadvertL2}, since high-degree vertices are handled by the faster and closer cache memory. 

\begin{figure*}[!htb]
    \centering
    \vspace{-0.1cm}
    \includegraphics[width=\textwidth]{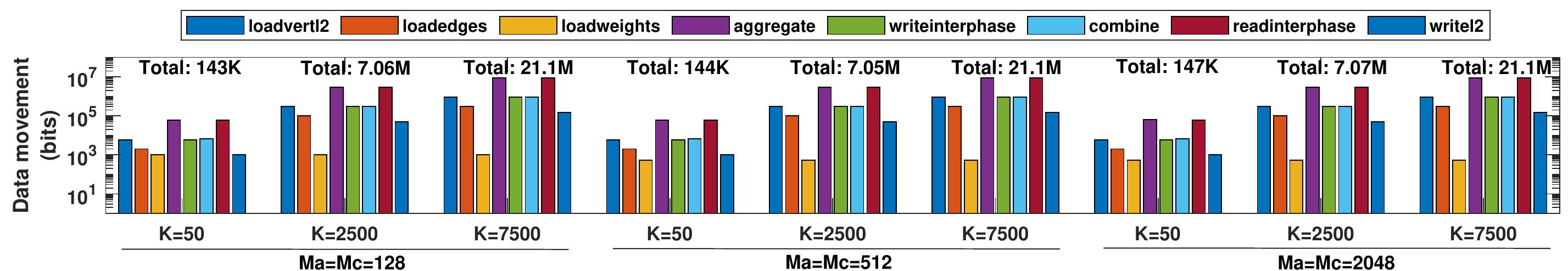}
    \vspace{-0.6cm}
    \caption{HyGCN amount of data movement with varying input graph and accelerator configuration.}
    \label{fig:HyGCN_movement_bar}
\end{figure*}

\begin{figure*}[!htb]
    \centering
    \vspace{-0.5cm}
    \includegraphics[width=0.8\textwidth]{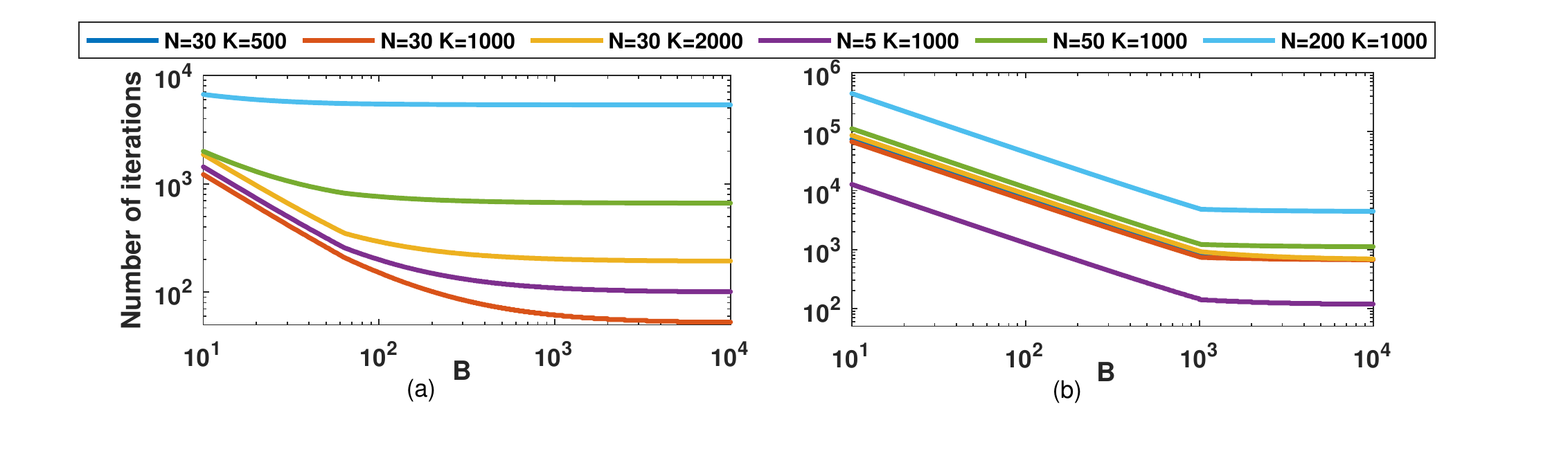}
    \vspace{-0.4cm}
    \caption{Number of iterations with varying memory bandwidth for different input workloads for (a) EnGN (b) HyGCN.} 
    \vspace{-0.6cm}
    \label{fig:iterations_MBW2_diffK_diffN}
\end{figure*}

Next, Fig. \ref{fig:iterations_MBW2_diffK_diffN}(a) shows how the total number of iterations scale with the memory bandwidth $B$, which is useful to detect bandwidth over-provisioning regions. We observe that bandwidth makes a bigger difference in smaller tiles as data is constantly moved in and out of the PE array. This is seen through both the saturation point and the value at which the curve saturates.


\begin{figure}[h]
    \centering
    \vspace{-0.4cm}
    \includegraphics[height=4cm, width=6cm]{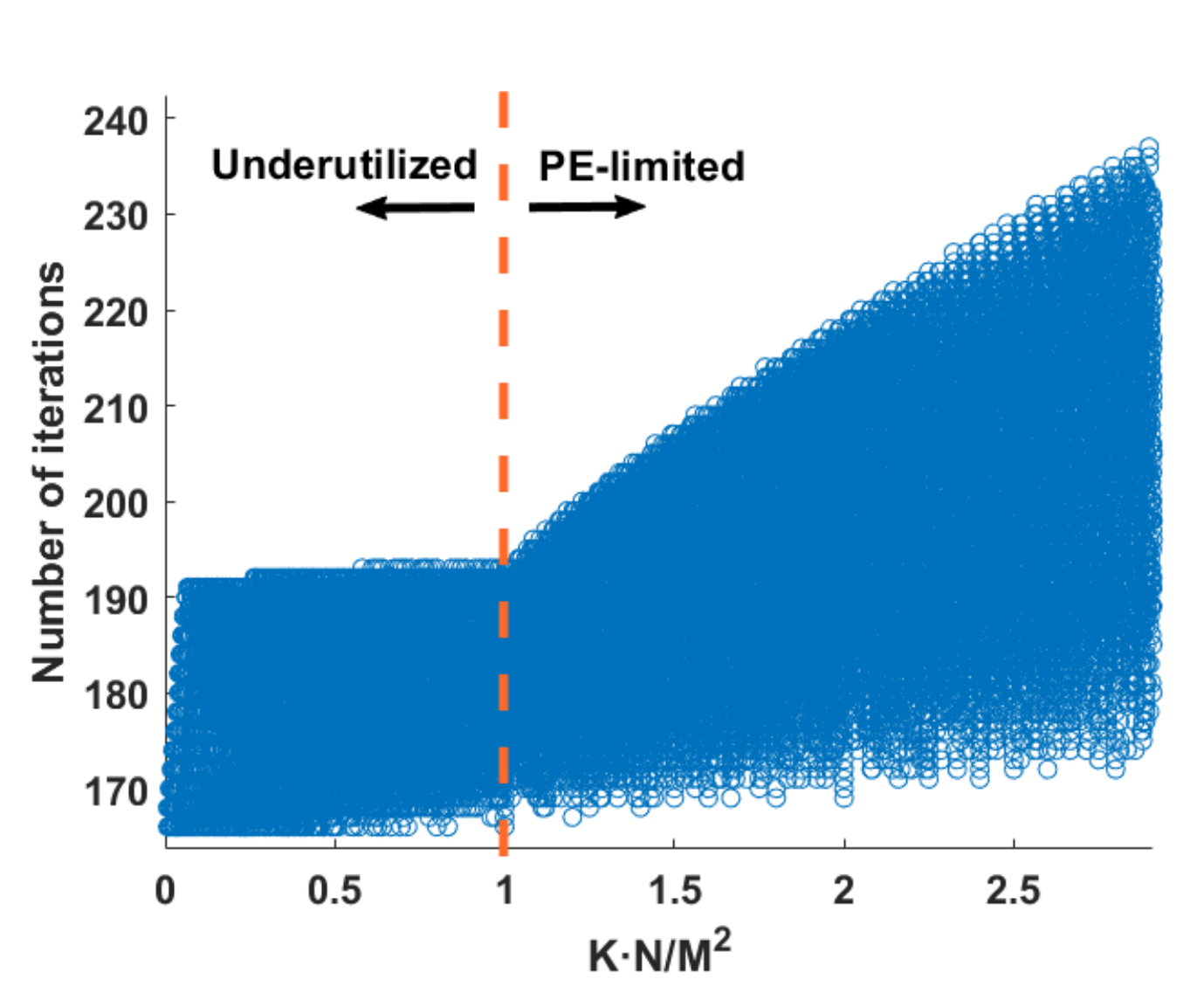}
    \vspace{-0.2cm}
    \caption{EnGN number of iterations of data movement with varying array fitting factor.}
    \vspace{-0.6cm}
    \label{fig:EnGN_iterations_KN/MM}
\end{figure}

We finally note that previous results, especially from Fig. \ref{fig:EnGN_bar}, seem to suggest that there is an optimal PE array size that depends on the tile size. This is because EnGN tries to fit the tiles in its PE array of size $M^{2}$ if $M=M'$, which is used both for aggregation and combination. Since a tile has a total size of $K$ nodes multiplied by the $N$ input features, an array fitting factor K$\cdot$N/M$^2$ has also been studied. 
By sweeping this factor as shown in Fig. \ref{fig:EnGN_iterations_KN/MM}, it is observed that whenever the number of elements to be placed in the array is large enough so that the number of PEs cannot host them, the number of iterations starts increasing since the aggregation and combination will take several steps to complete. This would help to explain why in Fig. \ref{fig:EnGN_bar} data movement first decreases and then increases with $M$, thus shows the potential of this methodology to reveal accelerator-specific behaviors.

\subsection{HyGCN}
Fig. \ref{fig:HyGCN_movement_bar} breaks down the total data movement of HyGCN. By comparing to Fig. \ref{fig:EnGN_bar}, we can provide interesting insights. First, we observe that data movement increases almost linearly with the tile size, but is independent of the array size. Second, HyGCN involves moves significantly more data than EnGN, due to its dual architecture and the need to write-read from the aggregation buffer. Moreover, HyGCN's aggregation depends on the size of the input feature vector $N$ instead of the output vector $T$, leading to large aggregations and phase transitions.

Further, Fig. \ref{fig:iterations_MBW2_diffK_diffN}(b) represents the HyGCN analogous to Fig. \ref{fig:iterations_MBW2_diffK_diffN}(a). The figure illustrates how HyGCN is also sensitive to bandwidth and how the saturation point, unlike in EnGN, is abrupt and independent on the tile size. This is mostly the effect of having aggregation buffers to orchestrate movement across phases, which makes HyGCN more bandwidth-hungry.



\begin{figure}[!tb]
    \centering
    \includegraphics[width=\columnwidth]{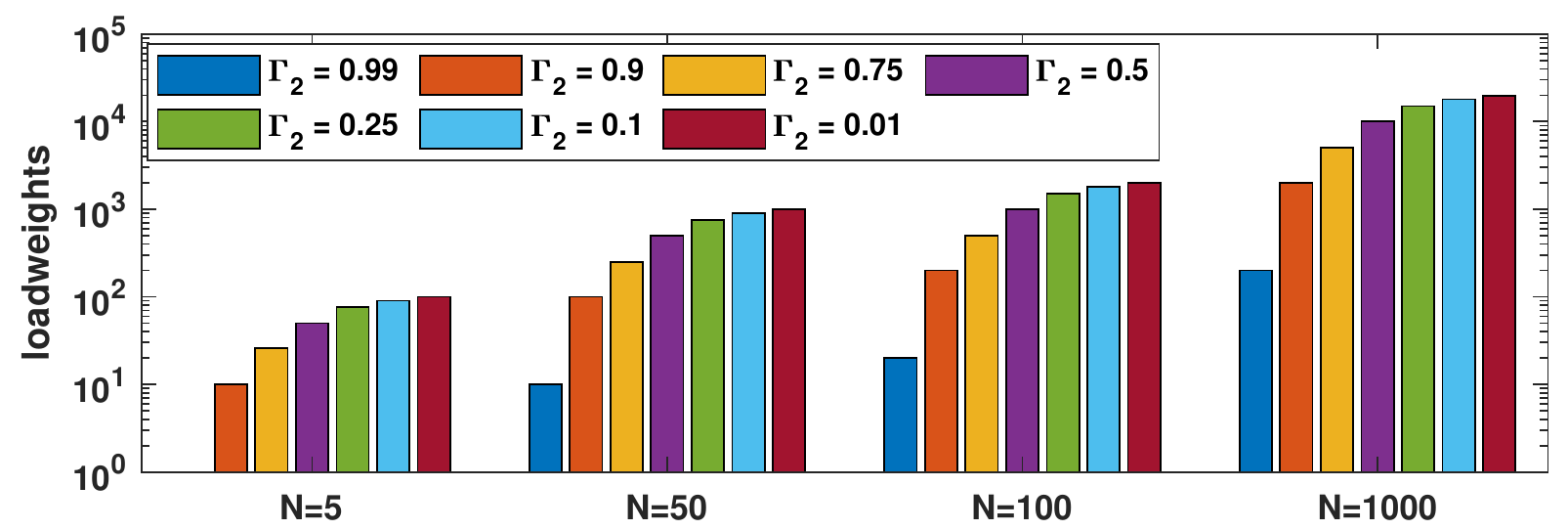}
    \vspace{-0.4cm}
    \caption{HyGCN \textit{loadweights} data movement with varying systolic array reuse.}
    \vspace{-0.3cm}
 \label{fig:HyGCN_weights_gamma}
\end{figure}

In order to model HyGCN-specific features, Fig. \ref{fig:HyGCN_weights_gamma} shows the effect of the systolic array reuse $\Gamma$ over \textit{loadweights}, for different graph depths $N$. We can see how large values of $\Gamma$, representing large reuse, alleviate the amount of weights to be loaded. This highlights the importance of tiling. 

\section{Conclusion} \label{cncl}
\vspace{-0.1cm}
We have presented a characterization method of the data movement within GNN accelerators based on analytical models. As a precursor of communication requirements, the proposed method can help comparing different architectures and explore their design spaces. In this work, we have tested the proposed method on two novel GNN accelerators and observed that (i) aggregation is accountable for a large fraction of data movement, (ii) the two architectures scale very differently, and (iii) memory limitations lead to an increase of total iterations, affecting latency. As future work, the analytical models for these and other accelerators will be expanded to model graph properties such as sparsity and validated against cycle-accurate simulations with dedicated tools.

\vspace{-0.2cm}
\section{Acknowledgements}
\vspace{-0.2cm}
The authors gratefully acknowledge support from the Spanish Ministry of Science and Innovation (MICINN) under grant EIN2019-103461 and from the European Comission under grant H2020-863337-WIPLASH.





\small{
\bibliographystyle{ieeetr}
\bibliography{ref}
}
\end{document}